\documentclass[preprint,superscriptaddress,nofootinbib]{revtex4}

\usepackage[T1]{fontenc}

\usepackage[english]{babel}
\usepackage{array,multirow} 
\usepackage{amsfonts,amsmath,amssymb}
\usepackage{graphicx,graphics}
\usepackage{color}
\usepackage[utf8]{inputenc}
\usepackage{caption}

\makeatother
\begin{document}

%\begin{frontmatter}

\title{Constraining the \texorpdfstring{$3-3-1$}{t331} model with heavy neutral leptons using \texorpdfstring{$(g-2)_{\mu}$}{mugm2} and dark matter observables}

%% Group authors per affiliation:
\author{C. E. Alvarez-Salazar}
\email{calvarez@ifi.unicamp.br, ORCID:0000-0003-3107-446X}
\affiliation{Instituto de F\'isica Gleb Wataghin - UNICAMP, 13083-859, Campinas,
SP, Brazil.}
%\author{C. E. Alvarez-Salazar}
%\ead{calvarez@ifi.unicamp.br, ORCID:0000-0003-3107-446X}
\author{O. L. G. Peres}
\email{orlando@ifi.unicamp.br, ORCID:0000-0003-2104-8460}
\affiliation{Instituto de F\'isica Gleb Wataghin - UNICAMP, 13083-859, Campinas,
SP, Brazil.}

%\author{O. L. G. Peres} 
%\ead{orlando@ifi.unicamp.br, ORCID:0000-0003-2104-8460}
%\address{Instituto de Física Gleb Wataghin-UNICAMP, 13083-859, Campinas/SP, Brazil}

\begin{abstract}

We find constraints on the highest scale of symmetry breaking of a model with gauge symmetry $SU(3)_C \otimes SU(3)_L \otimes U(1)_X$ with heavy neutral leptons in the fermion triplets, calculating the anomalous magnetic moment of the muon and using results of the relic abundance of dark matter and experiments searching for its direct detection.

In order to do this, we have calculated the one-loop contribution of new particles in the model to $(g-2)_{\mu}$, finding a favoured region for the scale at which $SU(3)_L$ is broken, and we have found lower bounds for this scale making a comparison of the predictions for the detection of a fermion dark matter candidate in the model in terms of simplified dark matter models, identifying the dominant portal for its interactions with standard model particles, and using constraints for the relic abundance and spin-independent scattering cross section of the fermion candidate with protons. 

\end{abstract}

\maketitle	

%\begin{keyword}
%Dark matter\sep Beyond the standard model\sep Simplified models \sep $3-3-1$ models
%\end{keyword}

%\end{frontmatter}

%\linenumbers

%%%%%%%%%%%%%%%%%%%%%%%%%%%%%%%%%%%%%%%%%%%%%%%%%%%%%%%%%%%%%%%%%%%%%%%%%
%%%%%%%%%%%%%%%%%%%%%%%% Introduction %%%%%%%%%%%%%%%%%%%%%%%%%%%%%%%%%%%
%%%%%%%%%%%%%%%%%%%%%%%%%%%%%%%%%%%%%%%%%%%%%%%%%%%%%%%%%%%%%%%%%%%%%%%%%

\section{Introduction}
Different experimental results point to the incompleteness of the standard model (SM) of particle physics as a theory describing the constituents of nature. The observation of neutrino oscillations, explained only in the framework of massive neutrinos \cite{Fukuda:1998mi}, the discrepancy between the SM prediction with the measured value of the anomalous magnetic moment of the muon \cite{Jegerlehner:2009ry}, the baryon asymmetry of the universe \cite{Cooke:2013cba} and the conclusion that only 5\% of the energy content of the universe is constituted by particles in the SM, are strong indications that physics beyond SM is needed.

The inclusion of matter with no interaction with light, explaining the reason to be called dark matter (DM), as a fundamental component of the theoretical model of particle physics has been considered mandatory in the last years due to different observations, both at galactic and cosmological scales~\cite{Bergstrom:2000pn,Bertone:2004pz}. Several results have led to the conclusion that 27\% of the energy in the universe correspond to matter in a form not included in the SM~\cite{Akrami:2018vks}.

Due to the unknown nature of DM, it is important to construct theoretical frameworks to describe its interactions with SM particles, and this could have different degrees of refinement. 

In the first place, the description of DM interactions in terms of effective field theories (EFT), constitutes a natural tool to perform model-independent analyses in terms of four-field operators for the interactions of DM with nucleons~\cite{Fitzpatrick:2012ix}, in the non-relativistic limit. One advantage of this treatment is the possibility to obtain stringent bounds on the physics scale suppressing higher dimensional operators~\cite{Abdallah:2015ter}.

The second possibility for a theory of DM-SM interactions includes the most important mediator states, leading to a better description of the kinematics of the interaction. This step forward is done in the so called simplified models~\cite{DiFranzo:2013vra,Berlin:2014tja}, where the interactions can be described in scalar or vector channels for DM particles of any spin. This treatment has been proved useful in the search for new physics, where the interactions of a small number of new particles give predictions for collider physics observables at the Large Hadron Collider (LHC)~\cite{Alves:2011wf}.

Finally, complete models not only include DM particles and the mediators of their interactions in their particle contents, but (sometimes) a plethora of new particles. These models could be considered extensions of the SM, and are inspired by the most diverse ideas~\cite{Roszkowski:2017nbc, Feng:2014uja}. Usually, these models are set to solve or explain issues of the SM, what leads to the appearance of particles with the required characteristics to be identified with DM.

In this work, we analyse a $SU(3)_C\otimes SU(3)_L\otimes U(1)_N$ framework ($3-3-1$ model, for short), in order to find constraints on the scale of $SU(3)_L$ symmetry breaking, which determines the mass scale of new particles in the model, using the measured values of the anomalous magnetic moment of the muon \cite{Bennett:2006fi}, the DM relic density \cite{Akrami:2018vks} and the exclusion limits set by DM direct detection experiments \cite{Aprile:2017iyp,Akerib:2018lyp}.

This paper is organized as follows. In section \ref{sec:simplifiedmodels} we will discuss briefly some motivations and characteristics of models beyond SM, emphasizing on simplified models for the description of DM-SM interactions. In section \ref{sec:fermionDM331} we present a summary of the $3-3-1$ model considered in this work, in order to find, in section \ref{sec:muong-2}, the contribution of new particles to the anomalous magnetic moment of the muon and, in section \ref{sec:domportal}, the dominant portal for DM-SM interactions. Furthermore, in section \ref{sec:constraintsDD} we find constraints on the mass of the mediator of DM-SM interactions, which can be translated in lower bounds for the $SU(3)_L$ symmetry breaking scale, and compare these results with the favoured window coming from the contributions to the anomalous magnetic moment of the muon. Finally, in section \ref{sec:summaryconstraints}, we make a comparison of our results with previous constraints found on the $3-3-1$ model, mainly based on LHC data, and present our conclusions in section \ref{sec:conclusions}.

%%%%%%%%%%%%%%%%%%%%%%%%%%%%%%%%%%%%%%%%%%%%%%%%%%%%%%%%%%%%%%%%%%%%%%%%%
%%%%%%%%%%%%%%%%%% Simplified Models %%%%%%%%%%%%%%%%%%%%%%%%%%%%%%%%%%%%
%%%%%%%%%%%%%%%%%%%%%%%%%%%%%%%%%%%%%%%%%%%%%%%%%%%%%%%%%%%%%%%%%%%%%%%%%

\section{Simplified models for the description of DM-SM interactions}\label{sec:simplifiedmodels}
Models going beyond the SM try to solve some of its problems or inconsistencies leading to different frameworks with its own structure~\cite{langacker2017standard}. For example, in order to solve the  gauge symmetry problem, associated with the chirality of electroweak interactions and the quantization of electric charge, a unification of interactions or a grand unified theory have been proposed~\cite{Buras:1977yy,Ellis:1982wr}; for the solution of the fermion problem, related to the existence of at least three lepton families with hierarchical masses, superstring theories~\cite{Schwarz:1982jn} or braneworld scenarios~\cite{Ida:1999ui} can give an explanation; the hierarchy problem, associated with divergent corrections to the Higgs boson mass, can be solved, for example, in the framework of supersymmetry~\cite{Martin:1997ns}, extended models~\cite{Schmaltz:2005ky,Han:2003wu}, dynamical mechanisms for symmetry breaking~\cite{Bardeen:1989ds} or large extra dimensions~\cite{Chacko:1999eb,ArkaniHamed:1999dz}.

Disregarding the details of any of these models, it is very desirable to have the possibility of embedding DM in their particle contents, which will interact with SM particles depending on the lagrangian of the model. These interactions are completely unknown at the moment, and can be described in terms of Simplified Models, where the mediator state is called ``portal''~\cite{Abdallah:2015ter,DiFranzo:2013vra}.
 
For models with a single candidate to DM, where a discrete symmetry protects the lightest odd particle of decaying, its interactions will depend on the particle types of the DM and the mediator. For example, in the case of fermionic DM, different from its own antiparticle and represented by a field $\psi$ interacting with SM particles through a scalar $S$ or vector $U_{\mu}$ portal, the interaction lagrangian can be written, respectively, as~\cite{Berlin:2014tja}
\begin{eqnarray}
    \mathcal{L}&=& g_{\psi}\overline{\psi}\psi S +\sum_f\frac{c_S m_{f}}{\sqrt{2}v_h}\overline{f}fS\label{eq:LpsiS},\\
    \mathcal{L}&=&g\overline{\psi}\gamma^{\mu}(V_{\psi}^U-A_{\psi}^U\gamma_5)\psi U_{\mu}+g\sum_f \overline{f}\gamma^{\mu}(V_{f}^U-A_{f}^U\gamma_5)f U_{\mu},\label{eq:LpsiU}
\end{eqnarray}
where $g_{\psi}$ and $g$ are couplings associated with the interaction of $\psi$ with $S$ and $U_{\mu}$, respectively, $c_S$ is a Yukawa-like coupling associated with the mass $m_f$ of the SM fermions $f$, $v_h$ is the vacuum expectation value of the Higgs boson, and $V_{\psi}^U$, $A_{\psi}^U$ ($V_{f}^U$, $A_{f}^U$), are vector and axial couplings of fermionic DM (SM) particles.

In order to identify DM particles, several direct and indirect experiments  have been performed, are in progress or under construction. In the case of indirect searches, the detection of the decay or annihilation products of DM particles is used as a probe for DM particles~\cite{slatyer2018indirect}. On the other hand, in direct detection experiments the scattering of DM particles could leave a signal in the detectors, which can be in the form of energy deposition, scintillation light or ionization~\cite{Klasen:2015uma}. 
A fundamental quantity in direct detection experiments is the spin-independent scattering cross section of a DM candidate with nucleons, which is usually obtained in these experiments and which gives the strongest constraints on DM observables. 

Using the interaction lagrangian in Eqs. (\ref{eq:LpsiS}) and (\ref{eq:LpsiU}), this spin-independent scattering cross section with protons, $\sigma_{p}^{\rm SI}$, for the cases of scalar and vector portals, are given by~\cite{Arcadi:2017kky}
\begin{equation}\label{eq:sigmaportals}
    \sigma_{p}^{\rm SI}=
    \begin{cases}
    \frac{\mu_{\psi p}^2}{\pi}\ g_{\psi}^2\ c_S^2 \frac{m_p^2}{v_h^2}\frac{f_N^2}{m_S^4}     & \text{Scalar portal},\\
    \frac{\mu_{\psi p}^2}{\pi}\frac{g^4}{m_U^4}\left(\frac{Zf_p+(A-Z)f_n}{A}\right)^2 & \text{Vector portal},
    \end{cases}
\end{equation}
where $\mu_{\psi p}$ is the reduced mass of the fermion DM-proton system, with masses $m_{\psi}$ and $m_p$, respectively, $f_N$ ($N=n,p$) is the effective coupling of DM with nucleons, $m_S$ and $m_U$ are the masses of the scalar and vector mediators, and the atomic and mass numbers of the target in a direct detection experiment are denoted by $Z$ and $A$, respectively.

%%%%%%%%%%%%%%%%%%%%%%%%%%%%%%%%%%%%%%%%%%%%%%%%%%%%%%%%%%%%%%%%%%%%%%%%%
%%%%%%%%%%%%%%%%%%%%%%% 3-3-1 Model %%%%%%%%%%%%%%%%%%%%%%%%%%%%%%%%%%%%
%%%%%%%%%%%%%%%%%%%%%%%%%%%%%%%%%%%%%%%%%%%%%%%%%%%%%%%%%%%%%%%%%%%%%%%%%

\section{The \texorpdfstring{$3-3-1$}{331} model with heavy neutral leptons}\label{sec:fermionDM331}

In this section we will present a review of a complete model beyond SM, for which a fermion DM candidate is contained in its particle spectrum, and for which we will calculate the contribution of new particles to the anomalous magnetic moment of the muon, and results on DM observables will be compared in terms of the simplified model predictions just discussed.

This extension corresponds to a model with gauge symmetry $SU(3)_C\otimes SU(3)_L \otimes U(1)_N$, which has been widely studied in the literature, due to its appealing characteristics. For example, the anomaly cancellation occurs only if the number of families is exactly three or one of its multiples~\cite{Sanchez-Vega:2016dwe}, the quantization of electric charge appears naturally~\cite{VanDong:2005ux}, neutrino masses can be included easily in the model, either by effective operators invariant under the gauge symmetry~\cite{Dong:2006mt} or by a double see-saw mechanism~\cite{Dias:2018ddy}, the strong CP problem can be solved and a nonthermal candidate for DM (an axion) can be included in its physical spectrum~\cite{Dias:2018ddy,dias2003naturally,Montero:2017yvy,romero2019constraints}, and the model is very interesting from the phenomenological point of view~\cite{Dong:2014wsa}. 

In this $3-3-1$ model, it is customary to define the electric charge operator as a linear combination of the diagonal generators ($T_i, \ I$) of the  group $SU(3)_L \otimes U(1)_N$ as $Q=T_3+\beta T_8+ N I$, where $\beta$ is an embedding parameter which determines the fermion assignments and electric charges of new particles.

Despite the multiple versions of $3-3-1$ models existing in the literature, in this work we will analyze a version which does not contain exotic quark charges, characterized by a parameter $\beta=1/\sqrt{3}$~\cite{Diaz:2004fs}. The reason for choosing this model in comparison with other versions lies on the fact that this model contains scalar, fermionic and vector DM candidates, but only the lightest of these particles can be considered a DM candidate, since all belong to the same discrete symmetry group used to stabilize DM \cite{Mizukoshi:2010ky}. 

In order to cancel gauge anomalies, we distribute the matter content of the model in the following way~\cite{Buras:2012dp}: the first two generations of left handed quarks transform as triplets, while the third generation transforms as an antitriplet, in the same way as left handed leptons: 
\begin{eqnarray}
q_{iL} & = & (u_i, d_i, J_i)^T_L \sim (3,3,0) \nonumber \\
q_{3L} & = & (d_3, -u_3, J_3)^T_L \sim (3,\bar{3},1/3)\\
F_{jL} & = & (l_j,\ \nu_j,\ E_j) \sim (1, \bar{3},-1/3)\nonumber,
\end{eqnarray}
with $i=1,2$ corresponding to the first and second quark generation, $j=~e,~\mu,~\tau$ denoting the three lepton families, and $\sim$ is used to indicate the transformation properties under the symmetry group. Note that the model contains one new up quark ($J_3$) and two new down quarks ($J_{1,2}$), alongside with three heavy leptons $E_j$. The lightest of these fermions can be made a DM candidate in the model using a discrete symmetry, as shown in \cite{Mizukoshi:2010ky}, and we will analyse the DM observables associated to the heavy lepton of the electron flavour in Secs. \ref{sec:domportal} and \ref{sec:constraintsDD}.

Right handed particles are singlets under $SU(3)_L$, with the following transformation rules:
\begin{eqnarray}
l_{jR} & \sim & (1,1,-1), \quad E_{jR}\sim (1,1,0) \quad (\rm{leptons}),\nonumber \\
u_{aR} & \sim & (3,1,2/3), \quad a=1, \dots ,4\quad (\textrm{up quarks})\\
d_{bR} & \sim & (3,1,-1/3), \quad b=1,\dots ,5 \quad (\textrm{down  quarks}),\nonumber
\end{eqnarray}
where the usual quark generations correspond to $a,b=1,\dots ,3$.

Symmetry breaking in the model happens through three scalar triplets $\eta$, $\rho$ and $\chi$, with the following components and transformation properties
\begin{small}
    \begin{equation}\label{eq:scalartriplets}
    \eta=\begin{pmatrix}
    \eta^0 \\ \eta^- \\ \eta^+
    \end{pmatrix} \sim \left(1,3,-\frac{2}{3}\right),
    \quad
    \rho = \begin{pmatrix}
    \rho^+ \\ \rho^0 \\ \rho^{\prime 0}  
    \end{pmatrix} \sim \left(1,3,\frac{1}{3}\right),
    \quad
    \chi = \begin{pmatrix}
    \chi^+ \\ \chi^{\prime 0} \\ \chi^0
    \end{pmatrix} \sim \left(1,3,\frac{1}{3}\right),
\end{equation}
\end{small}
interacting through a scalar potential consistent with renormalization and gauge invariance, on which a discrete symmetry $\mathbb{Z}_2$ is imposed in order to bring simplicity to the model and interpret the $\chi$ scalar triplet as the responsible for  breaking the $SU(3)_L$ symmetry to the SM one,
\begin{eqnarray}
\label{eq:Vscalar}
V\left(\eta,\rho,\chi\right) & = &  \mu_{1}^{2}\eta^{\dagger}\eta+\mu_{2}^{2}\rho^{\dagger}\rho+\mu_{3}^{2}\chi^{\dagger}\chi+\lambda_{1}\left(\eta^{\dagger}\eta\right)^{2}+\lambda_{2}\left(\rho^{\dagger}\rho\right)^{2}+\lambda_{3}\left(\chi^{\dagger}\chi\right)^{2} \nonumber\\
 &  & +\lambda_{4}\left(\chi^{\dagger}\chi\right)\left(\eta^{\dagger}\eta\right)+\lambda_{5}\left(\chi^{\dagger}\chi\right)\left(\rho^{\dagger}\rho\right)+\lambda_{6}\left(\eta^{\dagger}\eta\right)\left(\rho^{\dagger}\rho\right)\nonumber \\
  &  & +\lambda_{7}\left(\chi^{\dagger}\eta\right)\left(\eta^{\dagger}\chi\right)+\lambda_{8}\left(\chi^{\dagger}\rho\right)\left(\rho^{\dagger}\chi\right)+\lambda_{9}\left(\eta^{\dagger}\rho\right)\left(\rho^{\dagger}\eta\right)\nonumber\\
 &  & -\sqrt{2}f\epsilon_{ijk}\eta_{i}\rho_{j}\chi_{k}+\textrm{H.c.},
\end{eqnarray}
where $\mu_i$ ($i=1,2,3$) are quadratic self interactions that can be determined from the vacuum properties, $\lambda_i$ ($i=1,\dots, 9$) are quartic couplings determining the spectrum of scalars in the theory, and $f$ is a trilinear coupling usually taken proportional to the highest energy breaking scale in the model. The stability of this scalar potential was recently analyzed in \cite{Sanchez-Vega:2018qje}, where tree level constraints on the parameters of the model were obtained using copositivity conditions and current bounds on the masses of extra particles.

The scalar triplets in Eq.~\eqref{eq:scalartriplets} are responsible to give mass to all particles in the model. For example, physical scalars appear as the massive eigenstates of the mass matrices obtained when the scalar triplets get the vacuum expectation values
\begin{equation}\label{eq:scalarvevs}
    \langle \eta \rangle = \frac{1}{\sqrt{2}}\begin{pmatrix}v_{\eta}\\0\\0\end{pmatrix},\qquad
    \langle \rho\rangle = \frac{1}{\sqrt{2}}\begin{pmatrix}0\\ v_{\rho}\\ 0\end{pmatrix},\qquad
    \langle \chi \rangle = \frac{1}{\sqrt{2}}\begin{pmatrix}0\\0\\ v_{\chi}\end{pmatrix},
\end{equation}
and the components $\eta^0$, $\rho^0$ and $\chi^0$ are decomposed into their real ($R_{\eta,\rho,\chi}$) and imaginary ($I_{\eta,\rho,\chi}$) parts,
\begin{equation}
    \eta^0 = \frac{1}{\sqrt{2}}(R_{\eta}+iI_{\eta}),\quad 
    \rho^0 = \frac{1}{\sqrt{2}}(R_{\rho}+iI_{\rho}),\quad
    \chi^0 = \frac{1}{\sqrt{2}}(R_{\chi}+iI_{\chi}),
\end{equation}
leading to the mass matrices
\begin{equation}\label{eq:realmatrix}
    \mathcal{M}_R^2 = \begin{pmatrix}
2\lambda_1v_{\eta}^2+\frac{fv_{\rho}v_{\chi}}{v_{\eta}} & 
\lambda_6v_{\eta}v_{\rho}-fv_{\chi} & 
\lambda_4v_{\eta}v_{\chi}-fv_{\rho}\\
\lambda_6v_{\eta}v_{\rho}-fv_{\chi} & 
2\lambda_2v_{\rho}^2+\frac{fv_{\eta}v_{\chi}}{v_{\rho}} & 
\lambda_5v_{\rho}v_{\chi}-fv_{\eta}\\
\lambda_4v_{\eta}v_{\chi}-fv_{\rho} &
\lambda_5v_{\rho}v_{\chi}-fv_{\eta} &
2\lambda_3v_{\chi}^2+\frac{fv_{\eta}v_{\rho}}{v_{\chi}}
\end{pmatrix},
\end{equation}
\begin{equation}\label{eq:imagmatrix}
    \mathcal{M}_I^2 = \begin{pmatrix}
    \frac{fv_{\eta}v_{\chi}}{v_{\rho}} & fv_{\chi} & fv_{\eta}\\
    fv_{\chi} & \frac{fv_{\rho}v_{\chi}}{v_{\eta}} & fv_{\rho}\\
    fv_{\eta} & fv_{\rho} & \frac{fv_{\eta}v_{\rho}}{v_{\chi}}
    \end{pmatrix}.
\end{equation}

From the eigenvectors of the first of these matrices, three Higgs bosons $h$, $H_2$ and $H_3$ are obtained (from these three particles, we identify the lightest physical state with the SM Higgs boson), where approximate expressions for their masses, calculated using a perturbative approach, where found in~\cite{Alvarez-Salazar:2019asna}, and which we will use in this work in order to get a precise calculation of the masses of these particles. 

On the other hand, matrix~\eqref{eq:imagmatrix} gives a pseudo-scalar particle, denoted $H_0$, and two Goldstone bosons ($G_Z$ and $G_{Z^{\prime}}$, eaten by the neutral gauge bosons appearing in the physical spectrum, to be discussed later). 

To complete the spectrum of scalar states in the theory, the mass mixing matrices resulting from the charged symmetry states in \eqref{eq:scalartriplets} give two bosons, labelled  $H_W^{\pm}$ and $H_Y^{\pm}$, and the neutral states without a VEV in the same equation give an additional neutral scalar $H_V$ which can be another possible  candidate to DM in the model, stabilized by the same discrete symmetry than $E_e$.

The gauge sector of the model consists, besides the SM photon $A_{\mu}$ and the mediators of weak interactions $W_{\mu}^{\pm}$ and $Z_{\mu}$, on a charged gauge boson $Y_{\mu}^{\pm}$ and two additional neutral fields $V_{\mu}^0$ and $Z_{\mu}^{\prime}$. The first of these neutral fields can also be made a DM candidate in the model (under the same symmetry used to stabilize $E_e$ and $H_V$), and the second of these gauge fields appears from the $3\times 3$ mixing matrix of neutral gauge bosons, from which $A_{\mu}$ and $Z_{\mu}$ correspond to the other two eigenstates.

Complete expressions for the Yukawa lagrangian giving masses to fermions, particles in the Higgs sector, masses of gauge bosons and the expressions for the interactions between gauge bosons and fermions, can be found in~\cite{Buras:2012dp,Cao:2016uur}.

In the following section we will find the contribution of new particles in this $3-3-1$ model to the anomalous magnetic moment of the muon, in order to determine if the physical spectrum can give a sizeable deviation of the SM prediction of this precisely measured quantity, which can be used as a sensitive probe for models beyond SM.

%%%%%%%%%%%%%%%%%%%%%%%%%%%%%%%%%%%%%%%%%%%%%%%%%%%%%%%%%%%%%%%%%%%%%%%%%
%%%%%%%%%%%%%%%%%%%%%%%%%% Muon g-2 %%%%%%%%%%%%%%%%%%%%%%%%%%%%%%%%%%%%%
%%%%%%%%%%%%%%%%%%%%%%%%%%%%%%%%%%%%%%%%%%%%%%%%%%%%%%%%%%%%%%%%%%%%%%%%%

\section{Constraints from the muon anomalous magnetic moment}\label{sec:muong-2}
In order to set constraints on the scale of symmetry breaking of the $SU(3)_L$ group, we have calculated the contribution of new particles in the $3-3-1$ model to the anomalous magnetic moment of the muon, defined as 
\begin{equation}\label{eq:deltaamu}
    a_{\mu}=\frac{g_{\mu}-2}{2},
\end{equation}
where $g_{\mu}$ is the gyromagnetic ratio (or g-factor), in terms of which the orbital magnetic moment of the muon is written in terms of its spin as
\begin{equation}
    \vec{\mu} = -g_{\mu} \mu_0 \frac{\vec{\sigma}}{2},
\end{equation}
where $\mu_0$ is the Bohr magneton, and $\vec{\sigma}$ denotes the vector of Pauli matrices. 
The usefulness of this calculation lies on the fact that $\Delta a_{\mu}\equiv a_{\mu}^{\rm exp}-a_{\mu}^{\rm SM}$, where $a_{\mu}^{\rm exp}$ is the experimentally measured value and $a_{\mu}^{\rm SM}$ the SM prediction, is a quantity measured very precisely in particle physics~\cite{Bennett:2006fi}, and which can be used in order to set constraints on models beyond SM~\cite{Czarnecki:2001pv}.

\begin{figure}[!hbt]
    \centering
    \includegraphics[scale=0.75]{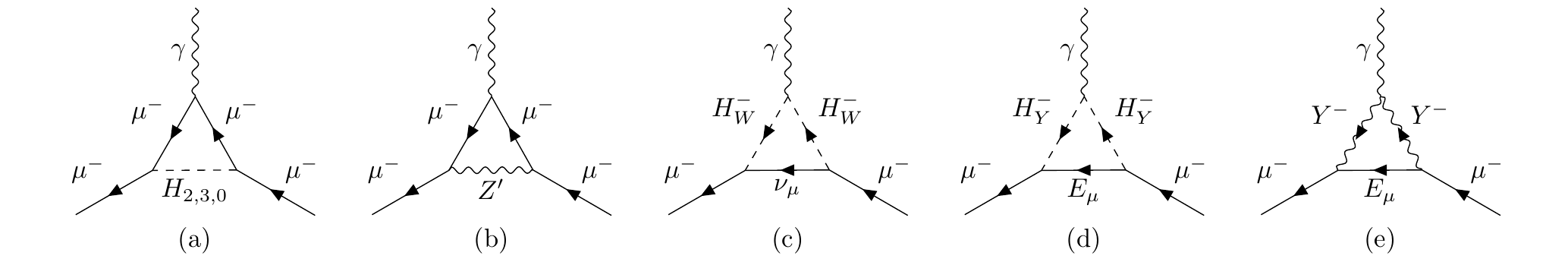}
    \caption{Contributions to the anomalous magnetic moment of the muon in the $3-3-1$ model considered in this work, for different exchanged particles (a) $H_2$, $H_3$ or $H_0$ neutral bosons, (b) $Z_{\mu}^{\prime}$ gauge boson, (c) $H_{W}^{\pm}$ and (d) $H_{Y}^{\pm}$ charged scalar bosons, respectively, and (e) ${Y}_{\mu}^{\pm}$ charged vector boson.}
    \label{fig:diagmuon}
\end{figure}

In order to calculate the contribution of new particles in the spectrum of the $3-3-1$ model considered in this work, we need to identify the possible one-loop lowest order type diagrams taking into account the interchange of both neutral and charged bosons. The contributing diagrams are shown in Fig.~\ref{fig:diagmuon}, and the general expressions for $\Delta a_{\mu}$ associated with diagrams of this kind can be found in \cite{Jegerlehner:2009ry}.

The contribution of diagrams characterized by the interchange of neutral scalar particles $H_2$, $H_3$ and $H_0$ (diagram (a) in Fig.~\ref{fig:diagmuon}) has the form
\begin{equation}\label{eq:scalardeltaamu}
    \Delta a_{\mu}^{S} = \frac{f_{S}^2}{8\pi^2}\lambda_S^2 \int_0^1 dx \frac{x^2(2-x)}{1-x+\lambda_{S}^2x^2},
\end{equation}
with $S=H_2,\ H_3,\ H_0$ a label indicating the particle interchanged (with mass $M_S$), and $\lambda_S = m_{\mu}/M_S$, with $m_{\mu}$ the muon mass.

In Eq.~\eqref{eq:scalardeltaamu}, $f_S$ ($S=H_2,\ H_3, H_0$), represents the vertex factor associated to the $\{\mu^-,\ \mu^-,\ S \}$ interaction given by
\begin{equation}
    f_{H_2} = \frac{U_{2,2}^H\ y_{\mu}}{\sqrt{2}}, \qquad 
    f_{H_3} = \frac{U_{2,3}^H\ y_{\mu}}{\sqrt{2}}, \qquad 
    f_{H_0} = \frac{U_{2,1}^h\ y_{\mu}}{\sqrt{2}},  
\end{equation}
where $y_{\mu}=\frac{m_{\mu}\sqrt{2}}{v_{\eta}}$ is the Yukawa coupling of the muon, and $U_{i,j}^H$ ($U_{i,j}^h$) is the $i,j$ element of the matrix transforming the $\{R_{\eta}, R_{\rho}, R_{\chi}\}$ ($\{I_{\eta}, I_{\rho}, I_{\chi}\}$) base to the mass eigenstates $\{h, H_2, H_3\}$ ($\{H_0, G_Z, G_{Z^{\prime}}\}$) (see the discussion on these physical states in Sec. \ref{sec:fermionDM331}), 
\begin{eqnarray}
    \begin{pmatrix} h \\ H_2 \\ H_3 \end{pmatrix} &=&  
\begin{pmatrix}
    U_{1,1}^H & U_{1,2}^H & U_{1,3}^H \\
    U_{2,1}^H & U_{2,2}^H & U_{2,3}^H \\
    U_{3,1}^H & U_{3,2}^H & U_{3,3}^H 
    \end{pmatrix}
    \begin{pmatrix} R_{\eta} \\ R_{\rho} \\ R_{\chi} \end{pmatrix}, \nonumber\\
    \begin{pmatrix} H_0 \\ G_Z \\ G_{Z^{\prime}} \end{pmatrix} &=&
    \begin{pmatrix}
    U_{1,1}^h& U_{1,2}^h & U_{1,3}^h \\
    U_{2,1}^h & U_{2,2}^h & U_{2,3}^h\\
    U_{3,1}^h & U_{3,2}^h & U_{3,3}^h 
    \end{pmatrix}
     \begin{pmatrix} I_{\eta} \\ I_{\rho} \\ I_{\chi} \end{pmatrix},
\end{eqnarray}
whose expressions, calculated using a pseudoinverse formulation of Rayleigh-Schrodinger perturbation theory were found in~\cite{Alvarez-Salazar:2019asna}. 

The contribution of the diagram where a $Z_{\mu}^{\prime}$ boson is interchanged (Fig. \ref{fig:diagmuon}(b)) has two contributions, due to the vector and axial couplings of this boson to the muon~\cite{Kelso:2013zfa}, and is given by
\begin{multline}\label{eq:vectorZPdeltaamu}
    \Delta a_{\mu}^{Z^{\prime}} = \frac{\lambda_{Z^{\prime}}^2}{4\pi^2}\left[
    (f_{Z^{\prime}}^V)^2\int_0^1 dx\frac{x^2(1-x)}{1-x+\lambda_{Z^{\prime}}^2x^2}\right.\\
    +\left.
        (f_{Z^{\prime}}^A)^2\int_0^1 dx\frac{x(1-x)(x-4)-2\lambda_{Z^{\prime}}^2x^3}{1-x+\lambda_{Z^{\prime}}^2x^2}
    \right],
\end{multline}
where $\lambda_{Z^{\prime}}=\frac{m_{\mu}}{M_{Z^{\prime}}}$, and $f_{Z^{\prime}}^{V,A}$ are the vector ($V$) and axial ($A$) couplings of the $Z_{\mu}^{\prime}$ to the muon, given by
\begin{eqnarray}
    f_{Z^{\prime}}^{V} &=&-\frac{e(-1+4s_W^2)}{c_W s_W\sqrt{3-4s_W^2}}, \nonumber\\
    f_{Z^{\prime}}^{A} &=&\frac{1}{-1+4s_W^2}f_{Z^{\prime}}^{V},
\end{eqnarray}
where $e$ is the electron charge, and $s_W$ and $c_W$ are the sine and cosine of the Weinberg angle, respectively.

The contribution of the diagram in Fig.~\ref{fig:diagmuon}(c), characterized by the interchange of a charged scalar $H_W^{\pm}$ is simpler than the others, due to the presence of the muon neutrino in the loop. In the limit $m_{\nu_{\mu}}\ll m_{\mu}$, the contribution is reduced to
\begin{equation}\label{eq:chargedHW}
    \Delta a_{\mu}^{H_W}=-\frac{f_{H_W}^2\lambda_{H_W}^2}{24\pi^2},
\end{equation}
where $f_{H_W}$ is the $\{\mu^-,H_W^-,\nu_{\mu}\}$ vertex factor, given by
\begin{equation}
    f_{H_W}=-\frac{m_{\mu}}{\sqrt{2 (v_{\eta}^2+v_{\rho}^2)}},
\end{equation}
and $\lambda_{H_W}=\frac{m_{\mu}}{M_{H_W}}$.

The interchange of charged scalars $H_Y^{\pm}$ in Fig.~\ref{fig:diagmuon}(d) gives the following contribution to the anomalous magnetic moment of the muon
\begin{multline}\label{eq:chargedHY}
    \Delta a_{\mu}^{H_Y}=-\frac{\lambda_{H_Y}^2}{8\pi^2}\left[
    (f_{H_Y}^S)^2\int_0^1 dx \frac{x(1-x)(x+\epsilon_{H_Y})}{(\epsilon_{H_Y}\lambda_{H_Y})^2(1-x)(1-\epsilon_{H_Y}^{-2}x)+x}\right. \\
    + \left.(f_{H_Y}^P)^2\int_0^1 dx \frac{x(1-x)(x-\epsilon_{H_Y})}{(\epsilon_{H_Y}\lambda_{H_Y})^2(1-x)(1-\epsilon_{H_Y}^{-2}x)+x}
    \right],
\end{multline}
where $\lambda_{H_Y}=\frac{m_{\mu}}{M_{H_Y}}$, $\epsilon_{H_Y}=\frac{m_{E_{\mu}}}{m_{\mu}}$, with $M_{H_Y}$ and $m_{E_{\mu}}$ the $H_Y^{\pm}$ and $E_{\mu}$ masses, respectively, and $f_{H_Y}^{S,P}$ are the scalar ($S$) and pseudoscalar ($P$) couplings of the $\{\mu^-,H_Y^-,E_{\mu}\}$ vertex, given by
\begin{eqnarray}
    f_{H_Y}^S &=& -\frac{1}{\sqrt{2(v_{\rho}^2+v_{\chi}^2)}}
    \left[m_{E_{\mu}}\frac{v_{\rho}}{v_{\chi}}+m_{\mu}\frac{v_{\chi}}{v_{\rho}}\right],\nonumber\\
    f_{H_Y}^P &=& \frac{1}{\sqrt{2(v_{\rho}^2+v_{\chi}^2)}}
    \left[m_{E_{\mu}}\frac{v_{\rho}}{v_{\chi}}-m_{\mu}\frac{v_{\chi}}{v_{\rho}}\right],
\end{eqnarray}

Finally, as the charged gauge boson $Y^{\pm}$ has vector and axial couplings (with the same strength) with $\mu$ and $E_{\mu}$, its contribution to the anomalous magnetic moment of the muon, calculated from the diagram in Fig. \ref{fig:diagmuon}(e), can be written as
\begin{equation}\label{eq:chargedY}
    \Delta a_{\mu}^{Y} =\frac{f_{Y}^2\lambda_{Y}^2}{4\pi^2}\int_0^1 dx
    \frac{2x^2(1+x)+\lambda_{Y}^2x(1-x)(x(1+\epsilon_Y^2)-2\epsilon_Y^2)}{(\epsilon_{Y}\lambda_{Y})^2(1-x)(1-\epsilon_{Y}^{-2}x)+x},
\end{equation}
with 
\begin{equation}
    f_Y=\frac{e}{2\sqrt{2}s_W}, \qquad \lambda_{Y}=\frac{m_{\mu}}{M_{Y}}\qquad \textrm{and}\qquad \epsilon_{Y}=\frac{m_{E_{\mu}}}{m_{\mu}},
\end{equation}
where $M_{Y}$ is the mass of the $Y^{\pm}$ boson.

\begin{figure}[htb!]
    \centering
    \includegraphics[scale=0.5]{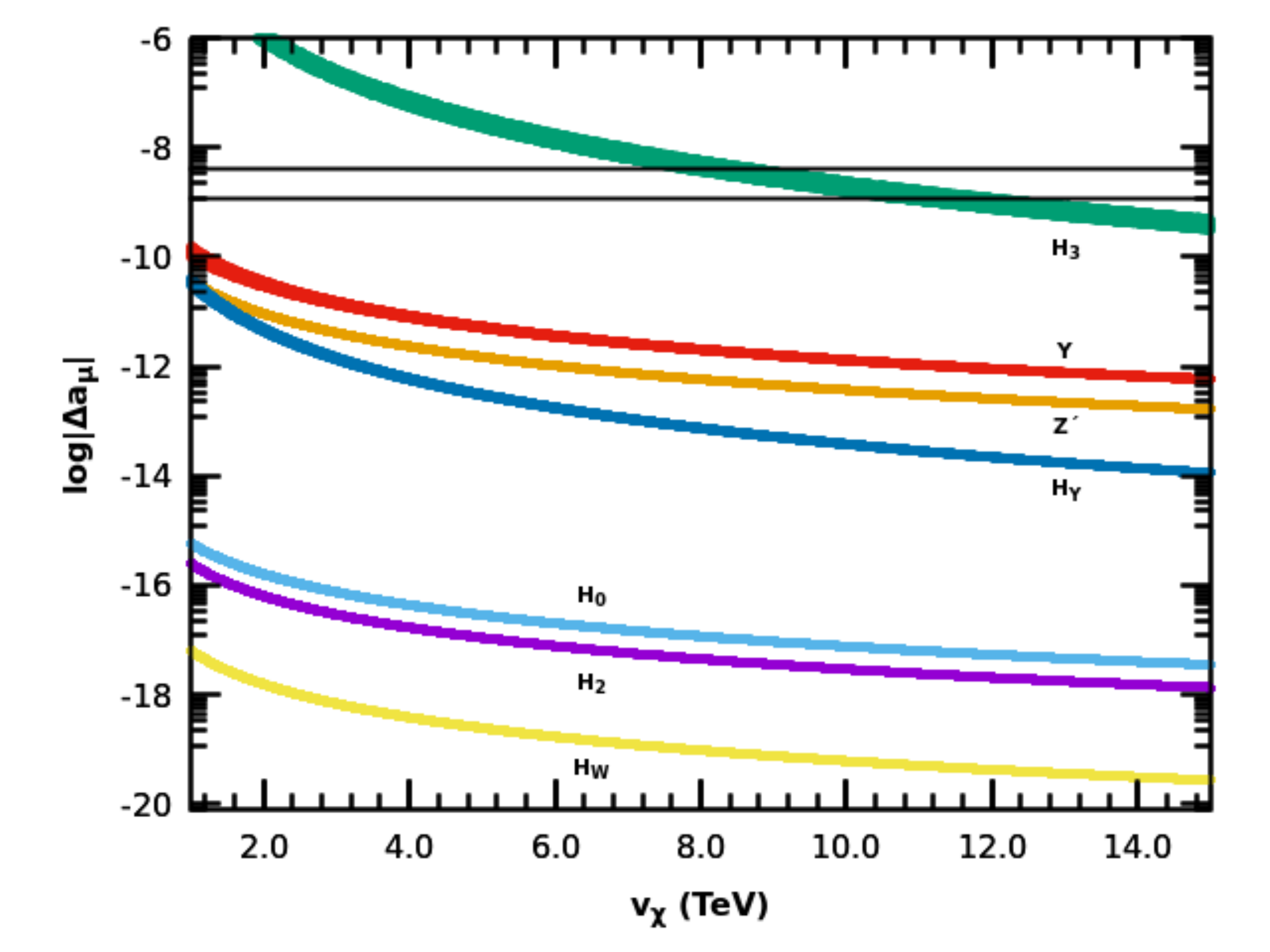}
    \caption{Contributions to $\Delta a_{\mu}$ of new particles in the $3-3-1$ model considered in this work. The contributions of $H_2$, $H_3$ and $Y_{\mu}^{\pm}$ are positive, and the contribution of all other particles is negative. The horizontal black lines represent the current value of $\Delta a_{\mu}$ (95\% C.L.)\cite{Tanabashi:2018oca}}
    \label{fig:gmuminus2}
\end{figure}

We have numerically calculated the contributions of all these new particles in the $3-3-1$ model (Eqs. \eqref{eq:scalardeltaamu} for $H_2$, $H_3$ and $H_0$, \eqref{eq:vectorZPdeltaamu} for $Z_{\mu}^{\prime}$,  \eqref{eq:chargedHW} for $H_W^{\pm}$, \eqref{eq:chargedHY} for $H_Y^{\pm}$ and \eqref{eq:chargedY} for $Y_{\mu}^{\pm}$) to the anomalous magnetic moment of the muon, and obtained the results shown in Fig. \ref{fig:gmuminus2}, where the contribution of each particle to $\Delta a_{\mu}$ is shown as a function of the $SU(3)_L$ symmetry breaking scale, on which the mass of each of these particles is strongly dependent. It is important to note here that the contributions of the CP-even scalars $H_2$ and $H_3$, and of the charged gauge boson $Y_{\mu}^{\pm}$ are positive, but the contributions of the CP-odd scalar $H_0$, the neutral gauge boson $Z_{\mu}^{\prime}$, and the charged scalars $H_W^{\pm}$ and $H_Y^{\pm}$ are negative.

In order to make a comparison with the reported value of $\Delta a_{\mu}$ \cite{Tanabashi:2018oca}, we have included in  Fig. \ref{fig:gmuminus2} the boundaries for this quantity (at 95\% C.L.), represented by  horizontal black lines. It is clear from this graph that the dominant contribution comes from the CP-even scalar $H_3$, which is at least two orders of magnitude greater than other contributions. From this graph it is possible to see that the $H_3$ contribution lies in the interval experimentally measured for values of the $SU(3)_L$ symmetry breaking scale such that
\begin{equation}\label{eq:boundsvchi}
7.2\ \textrm{TeV}\lesssim v_{\chi} \lesssim 12.2\ \textrm{TeV}\qquad (95\%\ \textrm{C.L}),
\end{equation}
determining a favoured window to look for the masses of the new particles present in the spectrum, as we will do in Sec. \ref{sec:constraintsDD} with the $Z_{\mu}^{\prime}$ boson.
%--------------- delta a_mu graph ------------------------------------

In the following section, we will discuss the results for DM observables in the $3-3-1$ model when we take the heavy lepton of the electron flavour as our DM candidate, identifying the dominant channel describing its interactions leading to a relic dark matter abundance consistent with the measurements performed by the Planck collaboration \cite{Aghanim:2018eyx}, and making a comparison of the results found in the complete $3-3-1$ model with the predictions of the simplified models presented in section~\ref{sec:simplifiedmodels}.

%%%%%%%%%%%%%%%%%%%%%%%%%%%%%%%%%%%%%%%%%%%%%%%%%%%%%%%%%%%%%%%%%%%%%%%%%
%%%%%%%%%%%%%%%%%%%%%%%%%% Dominant portal %%%%%%%%%%%%%%%%%%%%%%%%%%%%%%
%%%%%%%%%%%%%%%%%%%%%%%%%%%%%%%%%%%%%%%%%%%%%%%%%%%%%%%%%%%%%%%%%%%%%%%%%

\section{Identification of the dominant portal of DM-SM interactions}\label{sec:domportal}
In order to make a comparison of DM observables in the $3-3-1$ framework with the predictions of minimal DM models, we need to find the terms in the lagrangian with the structure presented in Eqs.~\eqref{eq:LpsiS} and~\eqref{eq:LpsiU}. Taking a look at the full lagrangian in the FeynRules~\cite{Christensen:2008py,Alloul:2013bka} implementation of the $\beta=1/\sqrt{3}$ version of the model~\cite{Cao:2016uur} and using the CalcHEP~\cite{Belyaev:2012qa} package, we have found that, in the case where the lightest odd particle under the discrete symmetry corresponds to the heavy fermion of the electron flavour, $E_e$, this particle interacts with four scalars in the physical spectrum of the model, and with two gauge bosons. 

The physical scalars interacting with $E_e$ are the eigenstates $h$, $H_2$ and $H_3$ of the real mass matrix \eqref{eq:realmatrix}, and the CP-odd state $H_0$ corresponding to the massive eigenstate of the imaginary mass matrix \eqref{eq:imagmatrix}.

On the other hand, $E_e$ interacts with two vector particles in the  model, the SM $Z_{\mu}$ boson and its heavier partner $Z^{\prime}_{\mu}$. This new gauge boson has a mass depending directly on the $SU(3)_L$ symmetry breaking scale $v_{\chi}$, 
\begin{equation}\label{eq:Mzpvchi}
    M_{Z^{\prime}}^2=\frac{g_W^2}{3-4s_W^2}\left(\frac{v_{\rho}^2(c_W^2-s_W^2)^2}{4c_W^2}+\frac{v_{\eta}^2}{4c_W^2}+c_W^2v_{\chi}^2\right),
\end{equation}
where $g_W$ is the $SU(3)_L$ coupling constant. 

\begin{figure}[htb!]
    \centering
    \includegraphics[scale=1.3]{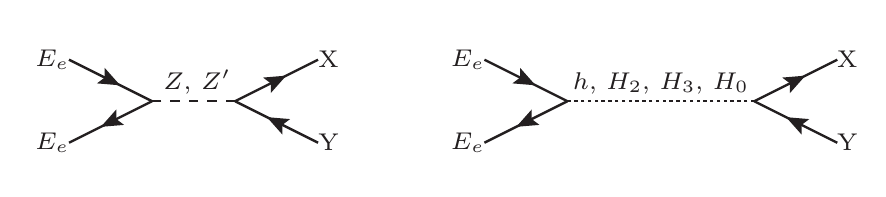}
   \caption{Feynman diagrams for pair processes contributing to the relic abundance of $E_1$}
    \label{fig:channels}
\end{figure}

In this way, we have seen that our DM candidate has the possibility to interact with SM particles both through scalar ($h$, $H_2$, $H_3$ and $H_0$) and vector ($Z_{\mu}$ and $Z_{\mu}^{\prime}$) channels, allowing us to identify the dominant portal describing the interactions of $E_e$, through the determination of resonances in the cross section for processes leading to the candidate abundance, $E_e\ \overline{E}_e\longleftrightarrow \rm{X}\ \rm{Y}$, as the ones shown in Fig.~\ref{fig:channels}. It has been shown~\cite{Escudero:2016gzx} that interactions of DM particles through SM portals, mediated by the Higgs particle $h$ or the $Z$ gauge boson are almost completely ruled out, and for this reason we don't analyze its interactions with our DM candidate.

\begin{figure}[hbt!]
  \centering
\includegraphics[scale=0.68]{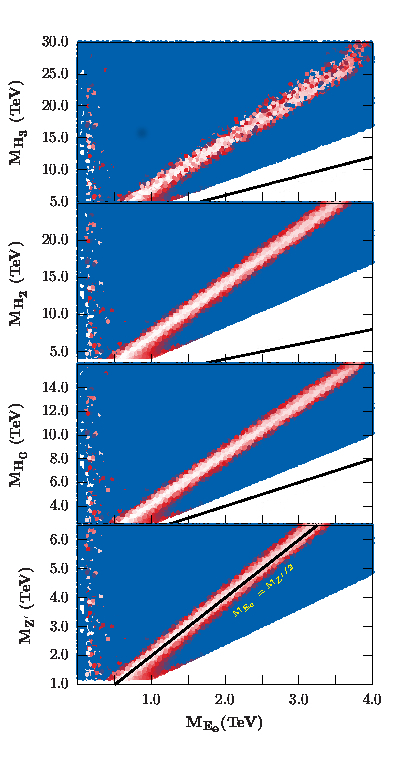}
  \caption{Identification of the dominant portal for the interactions of a fermion $E_e$ as candidate to DM in the $3-3-1$ model with $\beta=1/\sqrt{3}$. Resonances on the relic abundance
  are determined by the position of the two red lines present in the figures: if the solid black line, described by $M_{E_e}=M_{\rm Mediator}/2$ goes through the white stripe in the middle of the red lines giving a relic abundance consistent with Planck results~\cite{Aghanim:2018eyx}, the dominant portal is characterized by the interchange of the corresponding mediator.}\label{fig:resonances}
\end{figure}

Consequently, we have constructed the plots presented in Fig.~\ref{fig:resonances}, where we show a color map of the relic abundance of $E_e$ in the plane $M_{E_e}$ vs. $M_{\rm Mediator}$, where the color of each point indicates how the relic density $\Omega_{331}$ compares with the reported by the Planck collaboration $\Omega_{\rm Planck}$~\cite{Aghanim:2018eyx}: blue points correspond to 
$\Omega_{331} > \Omega_{\rm Planck}$, red points correspond (approximately) to $\Omega_{331} = \Omega_{\rm Planck}$, and the white dots in the middle of the two red lines in each diagram correspond to points where the relic density is very low or  approximately zero. These graphs have been produced using the micrOMEGAs package~\cite{Barducci:2016pcb}, using a modified version of the FeynRules~\cite{Christensen:2008py,Alloul:2013bka} implementation of the $3-3-1$ model in Ref.~\cite{Cao:2016uur}, including the calculation of the physical states $h$, $H_2$ and $H_3$ with the perturbative approach in~\cite{Alvarez-Salazar:2019asna}.

All plots in Fig.~\ref{fig:resonances} were obtained when the coupling parameters in the scalar potential~\eqref{eq:Vscalar} are of order 1 and we only guarantee that particles belonging to the same symmetry group that makes $E_e$ stable have greater mass. Nevertheless, this is not always possible, as can be seen from the white region in the lower right corner of each panel in Fig.~\ref{fig:resonances}, which corresponds to points in the parameter space where $E_e$ is no longer the lightest odd particle under the discrete symmetry, which is now substituted by the charged gauge boson $Y_{\mu}^{\pm}$.

It is important to note here that as the masses of $E_e$ and $Y_{\mu}^{\pm}$ become closer, annihilation involving $Y^{\pm}$ can change drastically the $E_e$ relic abundance~\cite{Griest:1990kh}. This degenerate regime is different to the general scenario depicted in Fig.~\ref{fig:resonances}, and its quantitative characteristics are not going to be described here, but our simulations for the scattering cross section with protons will give constraints for this regime.
Another point that is worth mentioning is the appearance of some points with a relic abundance consistent with the measured cosmological parameter $\Omega_{\rm Planck}$ on the left part of the graphs. These points correspond to the Higgs resonance on the scattering cross section, and it is always present in the model. Again, as Higgs mediated interactions are almost completely ruled out, we are not interested in their analysis in this work. 

So, from Fig.~\ref{fig:resonances}, we can see that the dominant portal for the interactions of $E_e$ is the $Z_{\mu}^{\prime}$ portal, as the red bands giving the relic abundance consistent with Planck results \cite{Aghanim:2018eyx} are placed symmetrically about the black line with equation $M_{E_e}=M_{Z^{\prime}}/2$, indicating a resonance for every value of $M_{Z^{\prime}}$ shown in the figure. This conclusion is consistent with the analysis of Ref.~\cite{Profumo:2013sca} which found constraints on the $Z_{\mu}^{\prime}$ boson mass using bounds obtained from direct detection experiments. 

%%%%%%%%%%%%%%%%%%%%%%%%%%%%%%%%%%%%%%%%%%%%%%%%%%%%%%%%%%%%%%%%%%%%%%%%%
%%%%%%%%%%%%%%%%%%%%%% Constraints %%%%%%%%%%%%%%%%%%%%%%%%%%%%%%%%%%%%%
%%%%%%%%%%%%%%%%%%%%%%%%%%%%%%%%%%%%%%%%%%%%%%%%%%%%%%%%%%%%%%%%%%%%%%%%%

\section{Constraints on the mass of the \texorpdfstring{$Z^{\prime}$}{zprime} boson and the \texorpdfstring{$v_{\chi}$}{vchi} VEV using DM direct detection experiments}\label{sec:constraintsDD}
Now that we have identified the dominant portal, we proceed to the determination of constraints on the properties of our DM candidate coming from the XENON1T direct detection experiment~\cite{Aprile:2017iyp}. In order to do so, we need to identify the parameters of the $3-3-1$ model corresponding to quantities entering in Eq.~\eqref{eq:sigmaportals}.

The terms in the lagrangian of the $3-3-1$ model which correspond to interactions of $E_e$ with the vector mediator $Z_{\mu}^{\prime}$, in the form given in Eq.~\eqref{eq:LpsiU}, can be written as
\begin{equation}\label{eq:L331Zp}
    \mathcal{L}_{331}\supset c_{1}\overline{E_e}\gamma^{\mu}(1-\gamma^5)E_e\  Z^{\prime}_{\mu}+c_{2}\overline{u}\gamma^{\mu}(c_3+\gamma^5)u\ Z^{\prime}_{\mu}+c_{4}\overline{d}\gamma^{\mu}(c_5+c_6\gamma^5)d\  Z^{\prime}_{\mu},
\end{equation}
where each $c_i$, $i=1,\dots,6$, are coefficients depending on the specific parameters of the model, and given by:
\begin{eqnarray}
c_1&=&\frac{e\left(1-s_W^2\right)}{2c_Ws_W\sqrt{3-4s_W^2}}, \nonumber\\
c_2&=&\frac{1}{2(1-s_W^2)}c_1,\nonumber\\
c_3&=&-1+\frac{8}{3}s_W^2,\\ 
c_4&=&-\frac{1}{3}c_2,\nonumber\\
c_5&=&2s_W^2+\left[\left(V^{\rm CKM}_{1,1}\right)^2+\left(V^{\rm CKM}_{2,1}\right)^2\right](3-4s_W^2)-\left(V^{\rm \rm CKM}_{3,1}\right)^2(3-2s_W^2),\nonumber\\
c_6&=&4s_W^2-c_5,\nonumber
\label{eq:parameters}
\end{eqnarray}
where $e$, $s_W$ and $c_W$ were defined in Sec. \ref{sec:muong-2}, and $V^{\rm CKM}_{i,j}$ are the $i,j$ components of the CKM matrix of the quark fields.

Taking these considerations into account, we can make the following identification of the parameters in~\eqref{eq:LpsiU}:
\begin{eqnarray}
%\begin{matrix}
 & g=c_1, & \\ 
V_{E_e}^{Z^{\prime}}=1, &  & A_{E_e}^{Z^{\prime}}=1,\\
V_{u}^{Z^{\prime}}=\frac{c_2c_3}{c_1}, & & A_{u}^{Z^{\prime}}=-\frac{c_2}{c_1}, \\
V_{d}^{Z^{\prime}}=\frac{c_4c_5}{c_1}, & &
A_{d}^{Z^{\prime}}=-\dfrac{c_4c_6}{c_1},\\
%\end{matrix}
\label{equivalence}
\end{eqnarray}
where we have replaced the indices for the DM candidate ($\psi$), the mediator ($U_{\mu}$) and the SM fermion ($f$) in Eq.~\eqref{eq:LpsiU} for the corresponding particle names in the $3-3-1$ model, $E_e$, $Z_{\mu}^{\prime}$ and $u, d$, respectively.

With this identification, we can proceed to the calculation of the spin-independent scattering cross section with protons, in order to set constraints on the masses of the DM candidate $E_e$ and the vector mediator $Z_{\mu}^{\prime}$. In order to do this, we have calculated $\sigma_p^{\rm SI}$ in two different situations: as given by Eq.~\eqref{eq:sigmaportals} (which we will call $\sigma_p^{Z^{\prime}}$ from now on), which assumes that $E_e$ interactions are mediated only by $Z_{\mu}^{\prime}$ and there are no other particles in the physical spectrum (besides SM particles), and considering the full particle content of the $3-3-1$ model (henceforth called $\sigma_p^{331}$), with all its possible portals (namely, $h$, $H_2$, $H_3$ and $H_0$), as done by the micrOMEGAs~\cite{Barducci:2016pcb} package.

When performing this calculation, we have found that the value of $\sigma_p^{Z^{\prime}}$ is always greater than $\sigma_p^{331}$ by a fixed factor of approximately $1.339$. This factor comes from neglecting the other contributing diagrams in the scattering cross section of $E_e$ with quarks, which produce a destructive interference with the diagram mediated by $Z_{\mu}^{\prime}$, in all the parameter space scanned in this work. So, in order to take into account this difference and be able to compare the results of the full $3-3-1$ model with the predictions of simplified models, we have included a normalization factor in 
Eq.~\eqref{eq:sigmaportals}\footnote{In our numerical calculations, we have found the ratio $\sigma_p^{331}/\sigma_p^{Z^{\prime}}=0.746846$, with a standard deviation $1.5\times 10^{-5}$, calculated over a sample with approximately $10^6$ data of $\sigma_p^{331}$ and  $\sigma_p^{Z^{\prime}}$, calculated simultaneously.}.

\begin{figure}[bht!]
    \centering
    \includegraphics[scale=0.85]{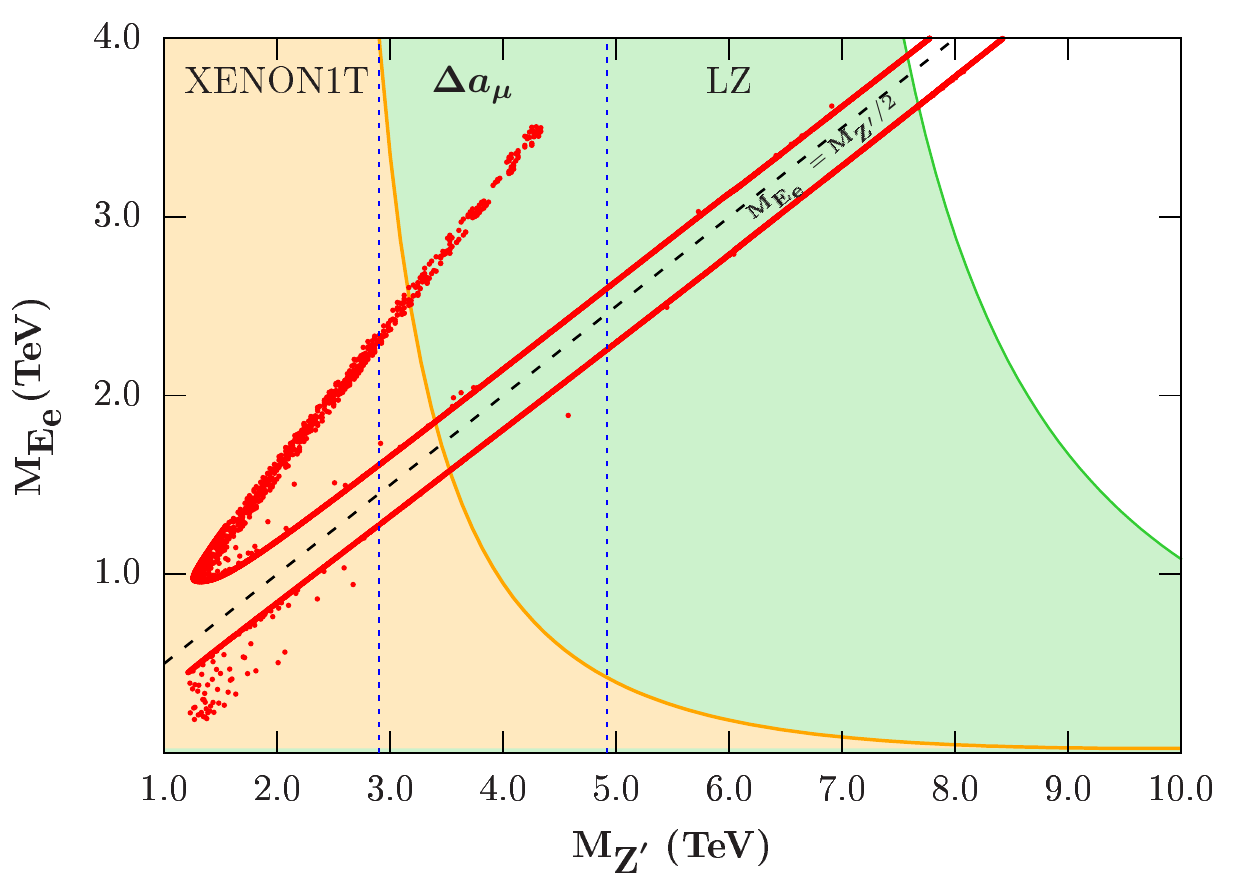}
    \caption{Constraints on $E_e$ and $Z^{\prime}$ masses set by XENON1T~\cite{Aprile:2017iyp}, and the future sensitivity of the LZ experiment~\cite{Akerib:2018lyp}. All red points have a relic abundance in the interval measured by the Planck collaboration~\cite{Aghanim:2018eyx}, and the vertical dashed blue lines correspond to the favoured region by the contribution of new particles to the anomalous magnetic moment of the muon.}
    \label{fig:constraints}
\end{figure}

In Fig.~\ref{fig:constraints} we show the constraints on the $E_e$ and $Z_{\mu}^{\prime}$ masses set by an extrapolation of the data from the XENON1T direct detection experiment~\cite{Aprile:2017iyp} (orange region), and the future WIMP sensitivity of the LZ experiment~\cite{Akerib:2018lyp} (green), where the red points have a relic density consistent with the measured by Planck~\cite{Aghanim:2018eyx}, and the vertical dashed blue lines correspond to the favoured region obtained from the contributions of new particles in the $3-3-1$ model to the anomalous magnetic moment of the muon, as shown in Fig. \ref{fig:gmuminus2}, with $M_{Z^{\prime}}$ calculated using Eq. \eqref{eq:Mzpvchi}.

From Fig. \ref{fig:constraints}, we can see that the measured value of DM relic abundance can be accomplished in two different regimes: the one dominated by the $Z^{\prime}$ resonance (two parallel bands symmetric about the dashed black line characterizing the resonance, $M_{E_e}=M_{Z^{\prime}}/2$), and the degenerate regime, when the mass of the DM candidate is very close to the mass of the $Y_{\mu}^{\pm}$ gauge boson in the same discrete group which makes $E_e$ stable. It is important to note here that this last regime requires some fine tuning in the parameters of the model.

Also from this Fig. we observe that, in order to make the $3-3-1$ model a suitable framework including fermion DM, there will be minimum values of $M_{E_e}$ and $M_{Z^{\prime}}$ which satisfy the constraints on the DM relic density and spin-independent scattering cross section with protons. Due to the direct dependence of the $Z_{\mu}^{\prime}$ mass on $v_{\chi}$ given by Eq. \eqref{eq:Mzpvchi}, the minimum values of the boson mass can be translated to constraints on the $SU(3)_L$ symmetry breaking scale, on which all masses of new particles are strongly dependent. The minimum values of the masses and the symmetry breaking scale, for each of the bands in Fig.~\ref{fig:constraints}, are shown in Table~\ref{tab:minimummasses}, which shows the exclusion limits set by the two different direct detection experiments considered.

\begin{table}[htb!]
    \centering
    \begin{tabular}{|c|p{4cm}|c|c|c|}
    \hline
& Regime & $M_{E_e}^{\rm min}$ (TeV) & $M_{Z^{\prime}}^{\rm min}$ (TeV) & $v_{\chi}^{\rm min}$ (TeV)\\
 \hline
 \multirow{3}{*}{\rotatebox{90}{ \tiny{XENON1T} }} 
 & Lower resonance & 1.6 & 3.6 & 8.8\\
 & Upper resonance & 1.9 & 3.4 & 8.4\\
& Degenerate region & 2.6 & 3.2 & 7.9\\
\hline
 \multirow{2}{*}{\rotatebox{90}{ \tiny{LZ} }} 
 & Lower resonance & 3.6 & 7.7 & 19\\
 & Upper resonance & 3.9 & 7.6 & 19\\
\hline
    \end{tabular}
    \caption{Minimum $Z_{\mu}^{\prime}$ and $E_e$ masses required for the $3-3-1$ model with $\beta=1/\sqrt{3}$ give a relic $E_e$ density consistent with Planck~\cite{Aghanim:2018eyx}, and evading the limits set by XENON1T~\cite{Aprile:2017iyp} and the future sensitivity of LZ~\cite{Akerib:2018lyp}.}
    \label{tab:minimummasses}
\end{table}

The minimum values of $M_{Z^{\prime}}$ are higher than the ones presented in Ref.~\cite{Salazar:2015gxa}, where the authors extend the analysis of decays of a new gauge boson to dilepton final states performed in~\cite{Aad:2014cka} to impose constraints on models with an additional particle of this type, in order to find lower bounds on the mass of the $Z_{\mu}^{\prime}$ boson of the $3-3-1$ model. On the other hand, our bounds on $M_{Z^{\prime}}$ are consistent with the results of Ref.~\cite{Queiroz:2016gif}, which analyze LHC data on flavor changing neutral currents (FCNC) and dilepton resonance searches.

Finally, the favoured window for the contribution of new particles in the model to $\Delta a_{\mu}$, giving the boundaries on $v_{\chi}$ shown in Eq. \eqref{eq:boundsvchi}, sets the following minimum and maximum values for the mass of the $Z_{\mu}^{\prime}$ gauge boson when we use Eq. \eqref{eq:Mzpvchi}:
\begin{equation}\label{eq:amuboundAP}
    2.9\ \textrm{TeV} \lesssim M_{Z^{\prime}} \lesssim 4.9\ \textrm{TeV}\ \qquad (95\%\ \textrm{C.L}),
\end{equation}
which, when combined with the constraints found with the results of the XE-NON1T \cite{Aprile:2017iyp} direct detection experiment gives a narrower window for the mass of this new gauge boson, with the same lower bounds shown in Table \ref{tab:minimummasses} and an upper bound given by the maximum value in Eq. \eqref{eq:amuboundAP}. On the other hand, the comparison of the favoured region represented by the vertical dashed blue lines in Fig. \ref{fig:constraints} with the constraint of the future sensitivity of the LZ experiment \cite{Akerib:2018lyp} points to the complete exclusion of the $3-3-1$ model with heavy neutral leptons as a suitable framework with a fermion candidate to DM.

%%%%%%%%%%%%%%%%%%%%%%%%%%%%%%%%%%%%%%%%%%%%%%%%%%%%%%%%%%%%%%%%%%%%%%%%%
%%%%%%%%%%%%%%%%%%% Summary of constraints %%%%%%%%%%%%%%%%%%%%%%%%%%%%%%
%%%%%%%%%%%%%%%%%%%%%%%%%%%%%%%%%%%%%%%%%%%%%%%%%%%%%%%%%%%%%%%%%%%%%%%%%

\section{Summary of constraints}\label{sec:summaryconstraints}
\begin{figure}[h]
    \centering
    \includegraphics[scale=0.82]{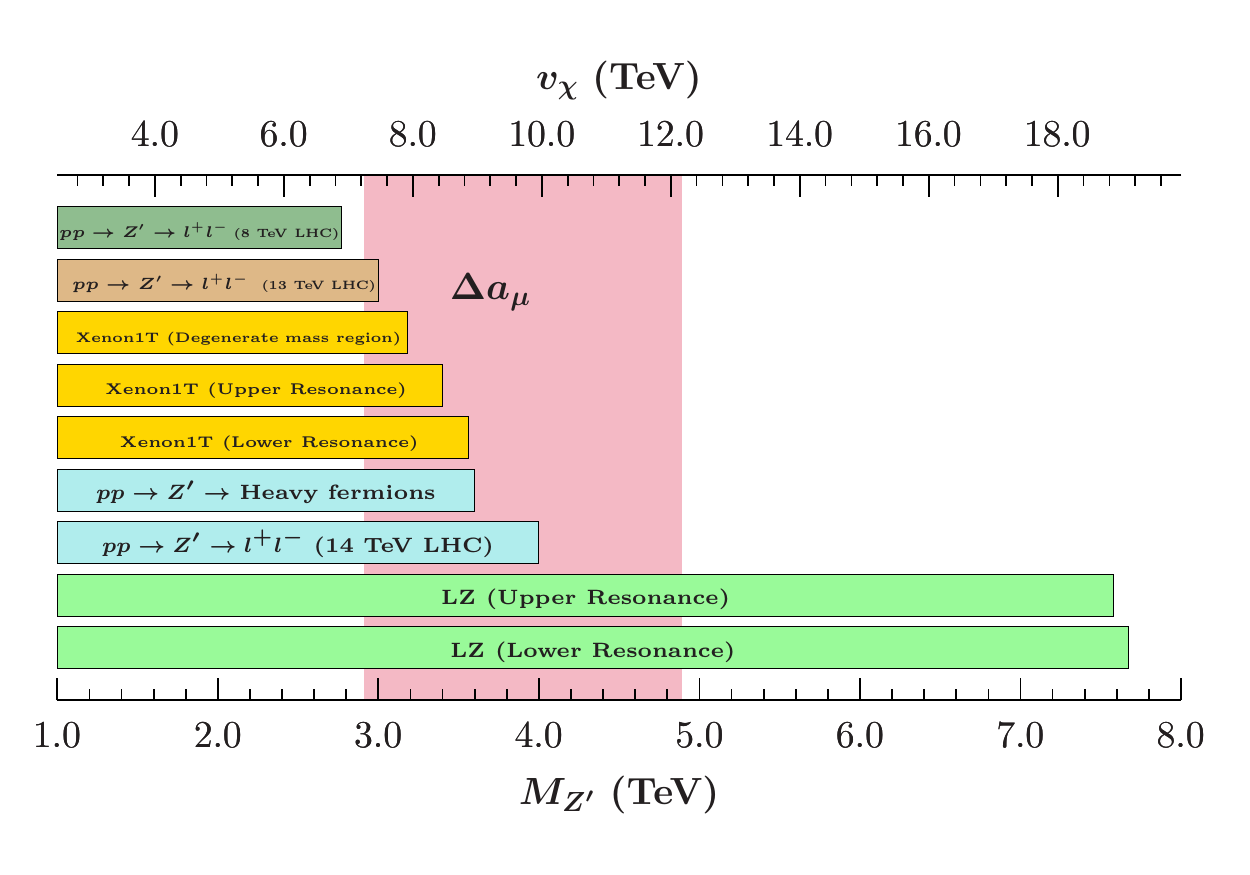}
    \caption{Excluded values for the $Z_{\mu}^{\prime}$ boson mass and the $v_{\chi}$ VEV set by different analyses: %the total cross section of 
    $Z^{\prime}$ decaying to dilepton final states in pp collisions using data from LHC  at a center of mass energy of 8 TeV
    (dark green)~\cite{Salazar:2015gxa} and 13 TeV (brown)~\cite{Queiroz:2016gif}, the exclusion limits set on this work (yellow bars) for the spin-independent cross section of fermion DM using XENON1T results~\cite{Aprile:2017iyp}, the negative results of dilepton searches at the 14 TeV LHC, both to heavy ($M\sim 1$ TeV) and SM fermions (cyan)~\cite{Cao:2016uur} and, finally, the constraints set on this work (light green) using the future sensitivity of the LZ experiment~\cite{Akerib:2018lyp}. The shaded vertical region shows the favoured values (95\% C.L) of $v_{\chi}$ and $M_{Z^{\prime}}$ obtained from the analysis of the contributions of new particles in the $3-3-1$ model with heavy neutral fermions to the anomalous magnetic moment of the muon.}
    \label{fig:constraintmZP}
\end{figure}

To summarize all constraints on the mass of the $Z_{\mu}^{\prime}$ boson, we have constructed Fig.~\ref{fig:constraintmZP}, where the horizontal bars indicate the excluded regions for $M_{Z^{\prime}}$ and $v_{\chi}$, related by the expression given in Eq.~\eqref{eq:Mzpvchi}, as obtained from different analyses. The shaded vertical region gives the favoured values (95\% C.L) of $v_{\chi}$ calculated from the total contribution of new physical states of the $3-3-1$ model considered in this work to the anomalous magnetic moment of the muon, $\Delta a_{\mu}$.

In this graph, the green and brown bars show the analyses of the process $pp\rightarrow Z^{\prime}\rightarrow l^+ l^-$ using LHC data at 8 ~\cite{Salazar:2015gxa} and 13 TeV~\cite{Queiroz:2016gif}, the yellow bars show the constraints set by the comparison of DM in the $3-3-1$ model with the predictions of simplified models, as performed in this work, the cyan bars, giving the current strongest constraints, show the bounds obtained by the decay of $Z^{\prime}$ to heavy and light fermions in the 14 TeV LHC~\cite{Cao:2016uur} and, finally, the light green bars show the bounds obtained from the future sensitivity of the LZ direct detection experiment \cite{Akerib:2018lyp}, obtained in this work.

It is important to remark here that the projected sensitivity of the LZ experiment will rule out $Z_{\mu}^{\prime}$ masses below 7.58 TeV, as shown in Table~\ref{tab:minimummasses}, and this result, combined with the favoured window for $\Delta a_{\mu}$ shown by the vertical shaded region in Fig. \ref{fig:constraintmZP}, will rule out the $3-3-1$ model with a  heavy neutral fermion as a candidate to DM as a suitable extension of the SM.

%%%%%%%%%%%%%%%%%%%%%%%%%%%%%%%%%%%%%%%%%%%%%%%%%%%%%%%%%%%%%%%%%%%%%%%%%
%%%%%%%%%%%%%%%%%%%%%% Conclusions %%%%%%%%%%%%%%%%%%%%%%%%%%%%%%%%%%%%%%
%%%%%%%%%%%%%%%%%%%%%%%%%%%%%%%%%%%%%%%%%%%%%%%%%%%%%%%%%%%%%%%%%%%%%%%%%

\section{Conclusions}\label{sec:conclusions}
In this work we have found constraints on a $3-3-1$ model, with a heavy neutral fermion as a DM candidate, coming from three different experimentally measured quantities: the anomalous magnetic moment of the muon, the DM relic density and the spin-independent scattering cross section of DM with protons.

In order to do this, we have calculated the contribution of new particles to the correction $\Delta a_{\mu}$ to the anomalous magnetic moment of the muon, finding a favoured window for the $SU(3)_L$ symmetry breaking scale of the model, namely $7.5\ \textrm{TeV}\lesssim v_{\chi} \lesssim 11\ \textrm{TeV}$. The importance of this symmetry breaking scale lies on the strong dependence of all masses of new particles in the model with this quantity.

On the other hand, considering a fermion DM candidate in the model with a relic density consistent with cosmological observations, and from the analysis of the spin-independent scattering cross section of this candidate with protons ($\sigma_p$), we were able to find minimum values of this symmetry breaking scale due to its relation with the mass of the vector portal of DM-SM interaction, a neutral gauge boson $Z_{\mu}^{\prime}$. 

In order to identify this particle as the dominant portal, we have made a comparison of the values of $\sigma_p^{331}$, the exact value of $\sigma_p$ considering all particles and interactions in the $3-3-1$ model, and $\sigma_p^{Z^{\prime}}$, the DM-proton scattering cross section calculated from a comparison of the $3-3-1$ model with the predictions of simplified models for DM interactions.

The comparison of the $3-3-1$ model predictions for the DM relic density and the spin-independent scattering cross section with the measurements of the Planck \cite{Aghanim:2018eyx} and XENON1T \cite{Aprile:2017iyp} collaborations, and the combination with the favoured window coming from the measurement of the anomalous magnetic moment of the muon \cite{Bennett:2006fi} lead to the bounds $v_{\chi}^{\rm min}\lesssim v_{\chi} \lesssim 11\ \textrm{TeV}$ and $M_{Z^{\prime}}^{\rm min}\lesssim M_{Z^{\prime}}  \lesssim 4.43\ \textrm{TeV}$ for the $SU(3)_L$ symmetry breaking scale and the $Z_{\mu}^{\prime}$ boson mass, respectively, where the values of $v_{\chi}^{\rm min}$ and $M_{Z^{\prime}}^{\rm min}$ depend on the regime associated to the production of fermion DM in the model, and are of order $8-9$ TeV and $3-4$ TeV, respectively.

Finally, the comparison of the favoured region for $\Delta a_{\mu}$ with the future sensitivity of the LZ direct detection experiment \cite{Akerib:2018lyp}, ruling out values of $v_{\chi}$ less than $\sim 19$ TeV and values of $M_{Z^{\prime}}$ lower than 7.5 TeV, leads to the conclusion that the $3-3-1$ model with heavy neutral fermions can not be a suitable extension of the SM when the DM candidate in the model corresponds to the heavy fermion of the electron flavour. In this case, other neutral particles being odd under the same discrete symmetry stabilizing DM, such as a scalar or a gauge boson in the physical spectrum, could give different results.

%%%%%%%%%%%%%%%%%%%%%%%%%%%%%%%%%%%%%%%%%%%%%%%%%%%%%%%%%%%%%%%%%%%%%%%%%
%%%%%%%%%%%%%%%%%%%%%% Acknowledgements %%%%%%%%%%%%%%%%%%%%%%%%%%%%%%%%%
%%%%%%%%%%%%%%%%%%%%%%%%%%%%%%%%%%%%%%%%%%%%%%%%%%%%%%%%%%%%%%%%%%%%%%%%%
%\section*{Acknowledgements}
\begin{acknowledgments}

C.E.A.S is grateful for the support of CNPq, under grant 159237/2015-7.
O.L.G.P. is  grateful for the support of FAPESP funding Grant  2014/19164-6, CNPq research fellowships 307269/2013-2 and 304715/2016-6 and for partial support from FAEPEX funding grant No. 2391/17. This study was financed in part by the Coordenação de Aperfeiçoamento de Pessoal de Nível Superior - Brasil (CAPES) - Finance Code 001.

\end{acknowledgments}

\bibliographystyle{apsrev4-1}
\bibliography{331SimplifiedModel}

\end{document}